# Controlling photonic spin Hall effect via exceptional points

Xinxing Zhou[1,2,*], Xiao Lin[1,*,†], Zhicheng Xiao[3], Tony Low[4], Andrea Alù[5,3], Baile Zhang[1,6,†], and Handong Sun[1,6,7,†]

[1]*Division of Physics and Applied Physics, School of Physical and Mathematical Sciences, Nanyang Technological University, Singapore 637371, Singapore*
[2]*Synergetic Innovation Center for Quantum Effects and Applications, School of Physics and Electronics, Hunan Normal University, Changsha 410081, China*
[3]*Department of Electrical and Computer Engineering, The University of Texas at Austin, Austin, TX 78712, USA*
[4]*Department of Electrical and Computer Engineering, University of Minnesota, Minneapolis, Minnesota 55455, USA*
[5]*Photonics Initiative, Advanced Science Research Center, City University of New York, New York, NY 10031, USA*
[6]*Centre for Disruptive Photonic Technologies (CDPT), School of Physical and Mathematical Sciences, Nanyang Technological University, Singapore 637371, Singapore*
[7]*MajuLab, CNRS-UCA-SU-NUS-NTU International Joint Research Unit, Singapore*
*These authors contributed equally to this work
[†]*Corresponding authors: xiaolinbnwj@ntu.edu.sg; blzhang@ntu.edu.sg; hdsun@ntu.edu.sg*

**The photonic spin Hall effect (SHE), featured by a spin-dependent transverse shift of an impinging optical beam driven by its polarization handedness, has many applications including precise metrology and spin-based nanophotonic devices. It is highly desirable to control and enhance the photonic SHE. However, such a goal remains elusive, due to the weak spin-orbit interaction of light, especially for systems with optical loss. Here we reveal a flexible way to modulate the photonic SHE via exceptional points, by exploiting the transverse shift in a parity-time (PT) symmetric system with balanced gain and loss. The underlying physics is associated with the near-zero value and abrupt phase jump of the reflection coefficients at exceptional points. We find that the transverse shift is zero at exceptional points, but it is largely enhanced in their vicinity. In addition, the transverse shift switches its sign across the exceptional point, resulting from spontaneous PT-symmetry breaking. Due to the sensitivity of transverse shift at exceptional points, our work also indicates that the photonic SHE can enable a precise way to probe the location of exceptional point in photonic systems.**

The photonic spin Hall effect (SHE), in analogy to SHE in electronic systems [1], is caused by the spin-orbit interaction of light at the interface, which is associated with the geometric Berry phase and originates from the transverse nature of light [2-7]. As schematically illustrated in Fig. 1(a), for an incident Gaussian beam, its different plane-wave components acquire different polarization rotations in the transverse direction upon reflection. The spatial-dependent polarization rotation directly leads to the spin splitting of the reflected beam in the transverse direction, featured by a spin-dependent transverse shif [8-17], where the spin state of photons corresponds to left-handed or right-handed circular polarizations (LCP or RCP). In other words, the photonic SHE introduces additional spin degrees of freedom for the flexible manipulation of light. As such, the photonic SHE shows promising applications in various realms, including optical sensing [18,19], spin-based nanophotonic devices [20], precise metrology [21,22], quantum information processing and plasmonics [6,23]. To enable and further extend these applications, a giant photonic SHE is desirable. However, the photonic SHE is generally weak, and its featured transverse shift is in the nanometer scale, due to weak spin-orbit interaction [4,24]. Therefore, controlling (e.g., enhancing) the photonic SHE in a flexible way remains an open challenge, highly desirable for its importance in modern photonics.

Several recent proposals to enhance the photonic SHE have been put forward, based on the Brewster effect [12,25], anisotropic and inhomogeneous metamaterials [20,26,27], and surface plasmon resonance phenomena [28,29]. These methods, however, are limited to transparent systems or systems with minimal material loss. Loss, however, is ubiquitous and oftentimes can be very large in optical systems, becoming largely detrimental to the enhancement of photonic SHE; a typical example of the effect of loss on the photonic SHE is shown in Figs. 1(c) and S2. Open questions are how to mitigate the influence of material loss on the photonic SHE and how to flexibly control it in systems with loss, especially for light with arbitrary polarization and incident angle.

Here we address these issues by introducing a viable scheme to flexibly control the photonic SHE in an optical system with loss, through the judicious incorporation and parametric modulation of exceptional points [30-34]. The exceptional point is a singularity in non-Hermitian systems, such as parity-time (PT) symmetric systems, in which at least two resonances coalesce in both eigenvalues and eigenfunction [35-39]. Although exceptional points have recently ignited tremendous interest [40-46], its connection with the photonic SHE has never been explored. We rely on the emerging new physics at the intersection of these two interesting yet distinct physical concepts. We find that the transverse shift stemming from the photonic SHE is zero at the exceptional points, but can be dramatically enhanced in the vicinity of exceptional points. Because of the spontaneous PT-symmetry breaking across the exceptional point, the giant transverse shifts have opposite signs at the two sides of exceptional points. Such a giant photonic SHE near the exceptional point can be flexibly designed to work with incident light of arbitrary polarization and incident angle, while this is hard to achieve in systems without exceptional points. Our work suggests that the PT-symmetric systems can be a promising and versatile platform to control the photonic SHE. Remarkably, due to the sensitivity of transverse shift at exceptional points, one can also precisely probe the location of exceptional points through the measurement of photonic SHE. Conversely, the photonic SHE can be an indispensable tool to study the exotic physics near/at the exceptional point in various photonic systems.

For the clarity of conceptual demonstration, here we consider the photonic SHE in a typical PT-symmetric structure, namely a bilayer structure with balanced loss and gain distribution as shown in Fig. 1(b). Each slab has a thickness of $d = \lambda/5$, where $\lambda$ is the wavelength in free space. Due to the scalability of Maxwell equations, the revealed phenomena below are applicable for an arbitrary value of $\lambda$. The slab with loss (gain) has a refractive index of $n_{\text{loss}} = n_R + in_I$ ($n_{\text{gain}} = n_R - in_I$), where $n_R = 3$ is chosen. Then $n_{\text{gain}} = n_{\text{loss}}^*$, which fulfills the PT-symmetry condition [34,38].

Due to the importance of exceptional points in the manipulation of photonics SHE, we briefly introduce the exceptional points in the studied structure in Fig. 1. For the two-port PT-symmetric structure in Fig. 1(b), the scattering matrix is $\bar{\bar{S}} = \begin{bmatrix} t & r_{+z} \\ r_{-z} & t \end{bmatrix}$ [47-49], where $t$ is the transmission coefficient and $r_{+z}$ ($r_{-z}$) is the reflection coefficient if the incident light propagates to the $+z$ ($-z$) direction [see supporting information]. For the scattering matrix $\bar{\bar{S}}$, its two eigenvalues are $\psi_\pm = t \pm \sqrt{r_{+z}r_{-z}}$. According to the generalized conservation relations, one has $|T - 1| = \sqrt{r_{+z}r_{+z}^* r_{-z} r_{-z}^*}$ [50], where $T = tt^*$ is the transmittance. Then, by using the generalized conservation relation in the expression of the two eigenvalues, we have $|\psi_+| = |\psi_-|$ if $T < 1$; this way, the PT-symmetric structure is in the exact PT symmetric phase. In contrast, if $T > 1$, $|\psi_+| \neq |\psi_-|$; the PT-symmetric structure exhibits the broken PT symmetric phase. Particularly, if $T = 1$ (which appears when $r_{+z}r_{-z} = 0$), the two eigenvalues, along with the two eigenfunctions for the scattering matrix $\bar{\bar{S}}$, coalesce to the same value; such a case is denoted as the exceptional point for the scattering matrix $\bar{\bar{S}}$.

The exceptional points of the scattering matrix $\bar{\bar{S}}$ are found to be dependent on the polarization and propagation direction of the incident beam. Due to the structural asymmetry of PT-symmetric systems, one generally has $r_{+z} \neq r_{-z}$, and thus there will be two different exceptional points, which appear at $r_{+z} = 0$ or $r_{-z} = 0$ (both cases have $T = 1$). Moreover, the reflection coefficient of *p*-polarized waves (i.e., TM waves or waves with horizontal polarization) is generally different from that of *s*-polarized waves (TE waves or waves with vertical polarization), namely $r^p \neq r^s$. When the propagation direction of the incident beam is known, there will also be two different exceptional points, which appear at $r^p = 0$ or $r^s = 0$. In short, for an incident beam with arbitrary propagation direction and linear polarization, there can be four different exceptional points, which appear at $r^p_{+z} = 0$, $r^p_{-z} = 0$, $r^s_{+z} = 0$, and $r^s_{-z} = 0$, respectively. To facilitate the discussion, we denote these four types of exceptional points as $EP^p_{+z}$, $EP^p_{-z}$, $EP^s_{+z}$, and $EP^s_{-z}$. Here and below, the superscripts *p*/*s* and the subscripts +z/−z indicate the polarization and propagation direction of incident light, respectively. These four types of exceptional points in the parameter space, and their dispersion as a function of the incident angle $\theta_i$ and the imaginary part of refractive index $|Im(n_{\text{loss/gain}})|$), are shown in Fig. 2.

We now proceed to calculate the photonic SHE near/at the exceptional points. We consider an incident Gaussian beam with linear polarization (*p*- or *s*-polarized); its angular spectrum is proportional to $e^{-w^2(k_{iy}^2 + k_{ix}^2)/4}$, where $w = 30\lambda$ is the beam waist and $k_{iy}$ ($k_{ix}$) is the $y$ ($x$) component of wave-vector. For the reflected (transmitted) waves, the spin-dependent transverse shift is dependent on the reflection (transmission) coefficients [12,21]. Due to reciprocity and structural asymmetry of the PT-symmetric system, if the incident angle is fixed, the transverse shift for reflected waves (transmitted waves) is sensitive to (independent of) the propagation direction of the incident beam. Therefore, we focus our discussion of photonic SHE on the reflected wave, since it provides an additional degree of freedom to control the photonic SHE compared with the transmitted wave.

For the analysis of spin-dependent transverse shifts, the reflected fields for LCP or RCP waves needs to be computed, so that their barycenter or centroid can be obtained. To be specific, a three-dimensional propagation model is needed for the derivation of the Gaussian beam reflection [see supporting information]. A Taylor series expansion for reflection coefficients, i.e., $r = r(\theta_i) + \frac{\partial r}{\partial \theta_i} \frac{k_{ix}}{k}$, should be introduced if $r(\theta_i) \to 0$, which arises at the exceptional point. Moreover, for the PT-symmetric structure in Fig. 1(b), it is noted that $\delta_{\text{LCP}} = -\delta_{\text{RCP}}$, where $\delta_{\text{LCP}}$ and $\delta_{\text{RCP}}$ are the transverse shifts of reflected waves with LCP and RCP, respectively. The study of the transverse shift of reflected waves with LCP, i.e., $\delta^p_{+z,\text{LCP}}$, $\delta^s_{+z,\text{LCP}}$, $\delta^p_{-z,\text{LCP}}$ and $\delta^s_{-z,\text{LCP}}$ in the parameter space in Fig. 2, can provide a full picture of the photonic SHE in the proposed PT-symmetric structure.

Figure 2 shows exotic photonic SHE near/at exceptional points. When the incident beam propagates towards the $+z$ direction in Fig. 2(a,b), the transverse shift $\delta_{+z,\text{LCP}}^{p/s}$ is suppressed down to zero at the exceptional point $\text{EP}_{+z}^{p/s}$. When the system slightly deviates away from $\text{EP}_{+z}^{p/s}$, the absolute value of transverse shift $|\delta_{+z,\text{LCP}}^{p/s}|$ becomes largely enhanced. To be specific, one has $|\delta_{+z,\text{LCP}}^{p}| > 10\lambda$ in Fig. 2(a), which is over two orders of magnitude larger than that found in typical air-glass interface (the related transverse shift is about $\lambda/10$) [4]. Moreover, the transverse shift $\delta_{+z,\text{LCP}}^{p/s}$ switches sign if the system moves across the $\text{EP}_{+z}^{p/s}$. This finding represents direct evidence of photonic SHE related to spontaneous PT-symmetry breaking. Similarly, when the incident beam propagates to the $-z$ direction in Fig. 2(c,d), exotic transverse shift $\delta_{-z,\text{LCP}}^{p/s}$ also arises near/at the exceptional point $\text{EP}_{-z}^{p/s}$.

We find that the switch of sign of transverse shifts across the exceptional point is dependent on the propagation direction of the incident beam. As a typical example, when $|Im(n_{\text{loss/gain}})|$ increases, the large-value $\delta_{-z,\text{LCP}}^{p}$ changes from positive to negative across $\text{EP}_{-z}^{p}$ as shown in Fig. 2(c), which is opposite to the sign change of $\delta_{+z,\text{LCP}}^{p}$ across $\text{EP}_{+z}^{p}$ in Fig. 2(a). Besides, it is noted that the PT-symmetric structure is in the broken (exact) PT-symmetric phase if the value of $|Im(n_{\text{loss/gain}})|$ is between (beyond) the two trajectories of $\text{EP}_{-z}^{p/s}$ and $\text{EP}_{+z}^{p/s}$ in Fig. 2. Therefore, it is reasonable to argue that the dependence of sign change of the giant transverse shift is related to the phase transition in PT-symmetric systems. On the other hand, the sign of the giant transverse shift across the exceptional point is also polarization dependent. For example, when $|Im(n_{\text{loss/gain}})|$ increases, the large $\delta_{+z,\text{LCP}}^{p}$ changes from negative to positive across $\text{EP}_{+z}^{p}$ for the $p$-polarized incident beam in Fig. 2(a); in contrast, the large $\delta_{+z,\text{LCP}}^{s}$ changes from positive to negative across $\text{EP}_{+z}^{s}$ for the $s$-polarized incident beam in Fig. 2(b).

For incident beams with arbitrary polarization and incident angle, the giant transverse shift can show up through judicious structural design, as shown in Figs. 2 and S4. This capability can be an advantage over other systems without exceptional points, which is generally dependent on the polarization or incident angle. Our findings in Fig. 2 thus indicate that the exceptional points allow for flexible parametric control of photonic SHE, including enhancing or suppressing the magnitude of the transverse shift and switching the sign of the transverse shift.

From the quantitative analysis of photonic SHE, the transverse shift is closely associated with the reflection coefficients [7,12]. Figure 3 shows the magnitude and phase of the reflection coefficients, i.e., $r_{+z}^{p}$, $r_{+z}^{s}$, $r_{-z}^{p}$, and $r_{-z}^{s}$. We find these exotic phenomena of photonic SHE near/at the exceptional points are caused by the near-zero value of the magnitude of reflection coefficients near the exceptional point in Fig. 3(a-d) and their abrupt phase change at the exceptional point in Fig. 3(e-h). Detailed explanation is given below.

Let us start with a *p*-polarized incident beam. The transverse shift is found to be proportional to $|r^p|^2[1 + \text{Re}\left(\frac{r^s}{r^p}\right)]$ [supporting information]. Since $|r^p| = 0$ at the exceptional point $EP^p$ in Fig. 3(a,b), the transverse shift is exactly zero at the exceptional point in Fig. 2(a,c). If $|r^p| \neq 0$, the magnitude of giant transverse shift becomes closely related to $|r^s|/|r^p|$, while the sign of the giant transverse shift is related to the term $\cos(\phi_s - \phi_p)$, where $\phi_{p/s}$ stands for the phase of $r^{p/s}$. Fig. 3(a-d) shows that there is a finite value of $|r^s|$ but a near-zero value of $|r^p|$ near $EP^p$. One would have a very significant value of $|r^s|/|r^p|$ near $EP^p$. This directly leads to the appearance of the giant photonic SHE near $EP^p$ for the *p*-polarized incident beam. Moreover, $r^p$ has an abrupt phase jump of $\pi$ at $EP^p$ in Fig. 3(e,f), while $r^s$ smoothly changes its phase at $EP^p$ in Fig. 3(g,h). Then if the system transits across the exceptional point $EP^p$, the term $\cos(\phi_s - \phi_p)$ will change its sign. This further gives rise to a sign switch in the giant transverse shift for the *p*-polarized incident beam. For the *s*-polarized incident beam, the transverse shift is proportional to $|r^s|^2[1 + \text{Re}\left(\frac{r^p}{r^s}\right)]$ [supporting information]. Due to the near-zero value and the abrupt phase jump of $r^s$ across the exceptional point $EP^s$ in Fig. 3(c,d,g,h), an analysis similar to the above one can also explain the exotic photonic SHE for the *s*-polarized incident beam in Fig. 2. Besides, the transverse shift near the exceptional points in Fig. 2 is not always large for arbitrary incident angles. The reason is that the transverse shift is also dependent on other terms, such as $\frac{\partial r^{s/p}}{\partial \theta_i}$, which may weaken the photonic SHE and play a dominant role over the enhancement term of $r^s/r^p$ or $r^p/r^s$.

We note that the exceptional point or the PT symmetry breaking can also appear in entirely passive optical systems. The reason is that the bilayer optical system in Fig. 1(b) with an arbitrary loss and gain profile can be mathematically transformed into a PT-symmetric one [40,51]. However, for passive optical systems, the reflection coefficients cannot be exactly zero at the exceptional points. Then it is reasonable to argue that the value of $r^s/r^p$ or $r^p/r^s$ at the exceptional points of passive optical systems will be much smaller than that at the exceptional points of PT-symmetric systems with balanced gain and loss distribution studied here. This way, according to the analysis of Fig. 3, a PT-symmetric system with balanced gain and loss distribution is needed to achieve a giant photonic SHE.

It is crucial to point out that the underlying physics for the flexible control of photonic SHE via exceptional point is not related to the Brewster effect. The explanation can be justified in two aspects. First, strictly speaking, the Brewster effect exists only at a specific interface, where the reflection coefficient at the Brewster angle is zero. Coincidentally, the zero-value reflection coefficients appear at the exceptional points in Fig. 3. However, they do not originate from the Brewster effect at a specific interface and indeed result from the interference effect in a two-layer slab. Second, for nonmagnetic systems, the Brewster effect has zero-value reflection coefficient only for *p*-polarized waves [52]. In contrast, in addition to the *p*-polarized waves, our designed non-magnetic PT-symmetric structure also has the zero-value reflection coefficient for the *s*-polarized waves at the exceptional points. As a result, in nonmagnetic systems, it is only possible to use the Brewster effect to enhance the photonic SHE for the *p*-polarized incident waves. As an advantage, our revealed mechanism can be applicable for both *p*-polarized and *s*-polarized incident waves.

To further the understanding of the exotic photonic SHE near/at exceptional points, Fig. 4 shows the field distribution of reflected beam with LCP, where a *p*-polarized Gaussian beam (i.e., an Hermitian-Gaussian beam with $TEM_{00}$ mode profile) with incident angle $\theta_i = 30°$ is chosen for illustration. On one hand, if the value of $|Im(n_{\text{loss/gain}})|$ increases, the transverse position of the beam centroid will transit from a positive (negative) value of $y$ to a negative (positive) one in Fig. 4(a,e) [Fig. 4(f,b)] when crossing the exceptional point. On the other hand, we find that, although the transverse spin splitting does not occur in Fig. 4(c,d), the reflected fields show a symmetric double-peak profile either along the $x$ direction at the exceptional point $EP^p_{-z}$ in Fig. 4(c) or along the $y$ direction at $EP^p_{+z}$ in Fig. 4(d); see more analysis in Figs. S4 and S5. Such a counterintuitive phenomenon emerges because the reflected fields in momentum space have $E^p_{\text{LCP}}(k_{rx}, k_{ry}) \propto \left(\frac{\partial r^p}{\partial \theta_i} k_{rx} + ik_{ry} r^s \cot \theta_i \right) e^{-w^2(k_{ry}^2 + k_{rx}^2)/4}$ at exceptional points, where $\theta_i = 30°$ here. To be specific, since $|\frac{\partial r^p}{\partial \theta_i}| \gg |r^s|$ at $EP^p_{-z}$ with $|Im(n_{\text{loss/gain}})| = 0.6043$ as shown in Fig. 4(g), the term related to $\frac{\partial r^p}{\partial \theta_i} k_{rx}$ in the momentum space plays a dominant role in determining the field distribution of the reflected beam in the real space. This way, the reflected beam has a symmetric double-peak profile along the $x$ direction in Fig. 4(c) and it is essentially a Hermitian-Gaussian beam with $TEM_{10}$ mode profile. In contrast, since $|\frac{\partial r^p}{\partial \theta_i}| \ll |r^s|$ at $EP^p_{+z}$ with $|Im(n_{\text{loss/gain}})| = 0.9425$ in Fig. 4(h), the term related to $ik_{ry} r^s \cot \theta_i$ becomes dominant. Then the reflected beam has a symmetric double-peak profile along the $y$ direction in Fig. 4(d) and becomes an Hermitian-Gaussian beam with $TEM_{01}$ mode profile. The reflected Hermitian-Gaussian beam with $TEM_{10}$ or $TEM_{01}$ mode profile revealed here has also been reported in systems without exceptional points [53]. Ref. [53] realizes it by tuning the waist of the incident Gaussian beam, where its incident angle is near the Brewster angle. As a distinctly different scheme, such a goal is achieved by merely letting the PT symmetric structure work at different exceptional points in Fig. 4(c,d).

Last but not least, there is an alternative definition of the scattering matrix for the two-port PT-symmetric structure in Fig. 1(b), namely $\bar{\bar{S}}' = \begin{bmatrix} r_{+z} & t \\ t & r_{-z} \end{bmatrix}$ [41,50], which corresponds to the permutation of the scattering matrix $\bar{\bar{S}} = \begin{bmatrix} t & r_{+z} \\ r_{-z} & t \end{bmatrix}$ [47]-[48]. Although both definitions of the scattering matrix are widely used, "because the permutation operation does not preserve the eigenvalues, these two different definitions of the $S$ (i.e., scattering) matrix lead to different criteria for the symmetric and broken-symmetry phases, as well as for the phase boundary (exceptional points)", as highlighted in Ref. [50]. Our results show that giant photonic SHE mainly happens near/at the exceptional point of the scattering matrix $\bar{\bar{S}}$, instead of the scattering matrix $\bar{\bar{S}}'$.

**Discussion**

In summary, we have revealed that the design of exceptional points in PT-symmetric systems can be a viable scheme to enhance and manipulate the photonic SHE. This scheme may mitigate the detrimental influence of material loss on the photonic SHE and may enable other promising applications, such as the design of novel optical sensors based on exceptional points and photonic SHE. In addition to the flexible control of photonic SHE studied here, we note that there are other promising applications and intriguing phenomena near/at exceptional points, such as the unidirectional invisibility [48], enhanced spontaneous emission rate [35] and sensing [45,46], and anomalous lasing [42,43]. Remarkably, these intriguing phenomena may have distinct features for different types of exceptional points. In addition to the second-order exceptional point studied here, other types of exceptional points have also been reported, including paired exceptional points [36], ring or line of exceptional points [54], exceptional surfaces [55], and high-order exceptional points [35,42,56,57]. Hence, our work may trigger other intriguing questions for photonic SHE, including the possibility to further enhance the photonic SHE via high-order exceptional points, and some new features of photonic SHE from PT-symmetric inhomogeneous and anisotropic metasurface where the geometric Pancharatnam-Berry phase [7] is involved (in our work, only the geometric Rytov-Vladimirskii-Berry phase is involved). This way, it is still highly desirable to continue the exploration of photonic SHE in other exotic systems with exceptional points or PT symmetry.

**Acknowledgment**

We thank Prof. Y. Chong for helpful discussions. This research was supported by the Singapore Ministry of Education through the Academic Research Fund [Tier 1-RG189/17-(s), Tier 1-RG105/16] and the National Natural Science Foundation of China (11604095).

**References**

[1] J. E. Hirsch, Phys. Rev. Lett. **83**, 1834 (1999).
[2] K. Y. Bliokh and Y. P. Bliokh, Phys. Lett. A **333**, 181 (2004).
[3] M. Onoda, S. Murakami, and N. Nagaosa, Phys. Rev. Lett. **93**, 083901 (2004).
[4] O. Hosten and P. Kwiat, Science **319**, 787 (2008).
[5] K. Y. Bliokh, A. Niv, V. Kleiner, and E. Hasman, Nature Photon. **2**, 748 (2008).
[6] K. Y. Bliokh, F. J. Rodríguez-Fortuño, F. Nori, and A. V. Zayats, Nature Photon. **9**, 796 (2015).
[7] X. Ling, X. Zhou, K. Huang, Y. Liu, C. W. Qiu, H. Luo, and S. Wen, Rep. Prog. Phys. **80**, 066401 (2017).
[8] A. Aiello, N. Lindlein, C. Marquardt, and G. Leuchs, Phys. Rev. Lett. **103**, 100401 (2009).
[9] Y. Qin, Y. Li, H. He, and Q. Gong, Opt. Lett. **34**, 2551 (2009).
[10] J.-M. Ménard, A. E. Mattacchione, M. Betz, and H. M. van Driel, Opt. Lett. **34**, 2312 (2009).
[11] M. Merano, N. Hermosa, J. P. Woerdman, and A. Aiello, Phys. Rev. A **82**, 023817 (2010).
[12] H. Luo, X. Zhou, W. Shu, S. Wen, and D. Fan, Phys. Rev. A **84**, 043806 (2011).
[13] N. Hermosa, A. M. Nugrowati, A. Aiello, and J. P. Woerdman, Opt. Lett. **36**, 3200 (2011).
[14] S. Goswami, M. Pal, A. Nandi, P. K. Panigrahi, and N. Ghosh, Opt. Lett. **39**, 6229 (2014).
[15] J. B. Götte, W. Löffler, and M. R. Dennis, Phys. Rev. Lett. **112**, 233901 (2014).
[16] W. J. M. Kort-Kamp, Phys. Rev. Lett. **119**, 147401 (2017).
[17] O. Takayama, J. Sukham, R. Malureanu, A. V. Lavrinenko, and G. Puentes, Opt. Lett. **43**, 4602 (2018).
[18] X. Zhou, L. Sheng, and X. Ling, Sci. Rep. **8**, 1221 (2018).
[19] L. Xie, X. Qiu, L. Luo, X. Liu, Z. Li, Z. Zhang, J. Du, and D. Wang, Appl. Phys. Lett. **111**, 191106 2017.
[20] N. Shitrit, I. Yulevich, E. Maguid, D. Ozeri, D. Veksler, V. Kleiner, and E. Hasman, Science **340**, 724 (2013).
[21] X. Zhou, Z. Xiao, H. Luo, and S. Wen, Phys. Rev. A **85**, 043809 (2012).
[22] X. Zhou, X. Ling, H. Luo, and S. Wen, Appl. Phys. Lett. **101**, 251602 (2012).
[23] K. Y. Bliokh, D. Smirnova, and F. Nori, Science **348**, 1448 (2015).
[24] L. Sun et al., Nature Photon. **13**, 180 (2019).
[25] L.-J. Kong, X.-L. Wang, S.-M. Li, Y. Li, J. Chen, B. Gu, and H.-T. Wang, Appl. Phys. Lett. **100**, 071109 (2012).
[26] X. Yin, Z. Ye, J. Rho, Y. Wang, and X. Zhang, Science **339**, 1405 (2013).
[27] P. V. Kapitanova et al., Nat. Commun. **5**, 3226 (2014).
[28] L. Salasnich, Phys. Rev. A **86**, 055801 (2012).
[29] X. Zhou and X. Ling, IEEE Photon. J. **8**, 4801108 (2016).
[30] C. M. Bender and S. Boettcher, Phys. Rev. Lett. **80**, 5243 (1998).
[31] M. V. Berry, D. H. J. O'Dell, J. Phys. Math. Gen. **31**, 2093 (1998).
[32] W. D. Heiss, J. Phys. A Math. Theor. **45**, 444016 (2012).
[33] L. Feng, R. E.-Ganainy, and L. Ge, Nat. Photon. **11**, 752 (2017).
[34] S. Longhi, EPL **120**, 64001 (2018).

[35] Z. Lin, A. Pick, M. Loncar, and A. W. Rodriguez, Phys. Rev. Lett. **117**, 107402 (2016).

[36] H. Zhou et al, Science **359**, 1009 (2018).

[37] R. E.-Ganainy, K. G. Makris, M. Khajavikhan, Z. H. Musslimani, S. Rotter, and D. N. Christodoulides, Nat. Phys. **14**, 11 (2018).

[38] M.-A. Miri and Andrea Alù, Science **363**, eaar7709 (2019).

[39] S. K. Özdemir, S. Rotter, F. Nori, and L. Yang, Nat. Mater. (2019).

[40] A. Guo, G. J. Salamo, D. Duchesne, R. Morandotti, M. V.-Ravat, V. Aimez, G. A. Siviloglou, and D. N. Christodoulides, Phys. Rev. Lett. **103**, 093902 (2009).

[41] Y. Chong, L. Ge, and A. Douglas Stone, Phys. Rev. Lett. **106**, 093902 (2011).

[42] H. Hodaei, M.-A. Miri, M. Heinrich, D. N. Christodoulides, M. Khajavikhan, Science **346**, 975 (2014).

[43] L. Feng, Z. J. Wong, R. M. Ma, Y. Wang, X. Zhang, Science **346**, 972 (2014).

[44] X. Lin, R. Li, F. Gao, E. Li, X. Zhang, B. Zhang, and H. Chen, Opt. Lett. **41**, 681 (2016).

[45] W. Chen, S. K. Özdemir, G. Zhao, J. Wiersig, L. Yang, Nature **548**, 192 (2017).

[46] H. Hodaei, A. U. Hassan, S. Wittek, H. G.-Gracia, R. E.-Ganainy, D. N. Christodoulides, and M. Khajavikhan, Nature **548**, 187 (2017).

[47] A. Mostafazadeh, Phys. Rev. Lett. **102**, 220402 (2009).

[48] Z. Lin, H. Ramezani, T. Eichelkraut, T. Kottos, H. Cao, D. N. Christodoulides, Phys. Rev. Lett. **106**, 213901 (2011).

[49] L. Feng, Y.-L. Xu, W. S. Fegadolli, M.-H. Lu, J. E. B. Oliveira, V. R. Almeida, Y.-F. Chen, and A. Scherer, Nat. Mater. **12**, 108 (2013).

[50] L. Ge, Y. Chong, and A. D. Stone, Phys. Rev. A **85**, 023802 (2012).

[51] C. E. Rüter, K. G. Makris, R. E. Ganainy, D. N. Christodoulides, M. Segev, and D. Kip, Nat. Phys. **6**, 192 (2010).

[52] X. Lin, Y. Shen, I. Kaminer, H. Chen, and M. Soljačić, Phys. Rev. A **94**, 023836 (2016).

[53] J.-L. Ren, B. Wang, Y.-F. Xiao, Q. Gong, and Y. Li, Appl. Phys. Lett. **107**, 111105 (2015).

[54] B. Zhen et al., Nature **525**, 354 (2015).

[55] H. Zhou, J. Y. Lee, S. Liu, and B. Zhen, Optica **6**, 190 (2019).

[56] G. Demange and E.-M. Graefe, J. Phys. A: Math. Theor. **45**, 025303 (2012).

[57] H. Jing, S. K. Özdemir, H. Lü, and F. Nori, Sci. Rep. **7**, 3386 (2017).

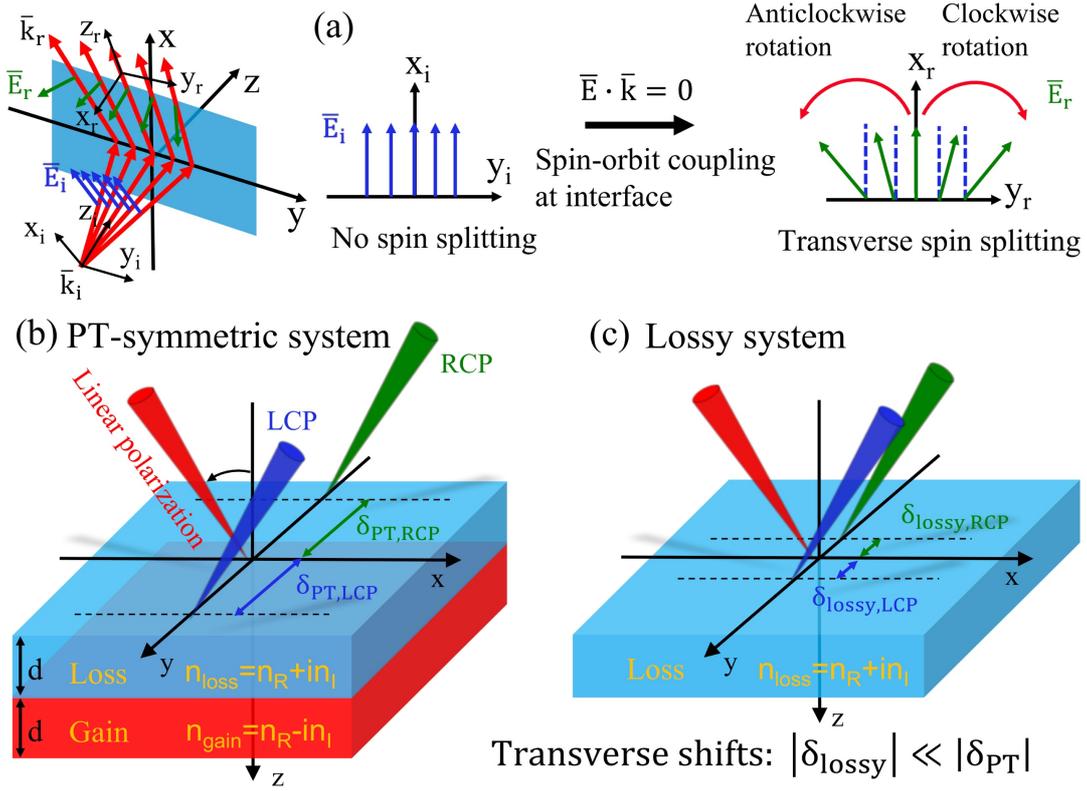

FIG. 1. Schematic of enhanced photonic spin Hall effect (SHE) via exceptional points. (**a**) Schematic illustration of photonic SHE for the reflected waves. The blue and green arrows indicate the directions of polarization vectors or the electric fields associated with each plane wave component before and after reflection. (**b**) Photonic SHE in a PT-symmetric system with the balanced gain and loss distribution. (**c**) Photonic SHE in a lossy system. Due to the photonic SHE, the reflected beam with left-handed or right-handed circular polarization (LCP or RCP) is split in the transverse or $y$ direction and have opposite transverse shifts, namely $\delta_{LCP} = -\delta_{RCP}$. For the reflected beam with LCP or RCP, the transverse shift near the exceptional point in PT-symmetric systems in (b) can be much larger than that in lossy systems in (c), i.e., $|\delta_{lossy}| \ll |\delta_{PT}|$; see Figs. 2 and S2. The slab with loss (gain) has a refractive index of $n_{loss} = n_R + in_I$ ($n_{gain} = n_{loss}^*$). The incident light is a Gaussian beam with horizontal (*p*-polarized waves) or vertical (*s*-polarized waves) polarization and has a beam waist of $30\lambda$. As conceptual demonstration, $n_R = 3$ is chosen, the two slabs with loss or gain have an identical thickness of $d = \lambda/5$.

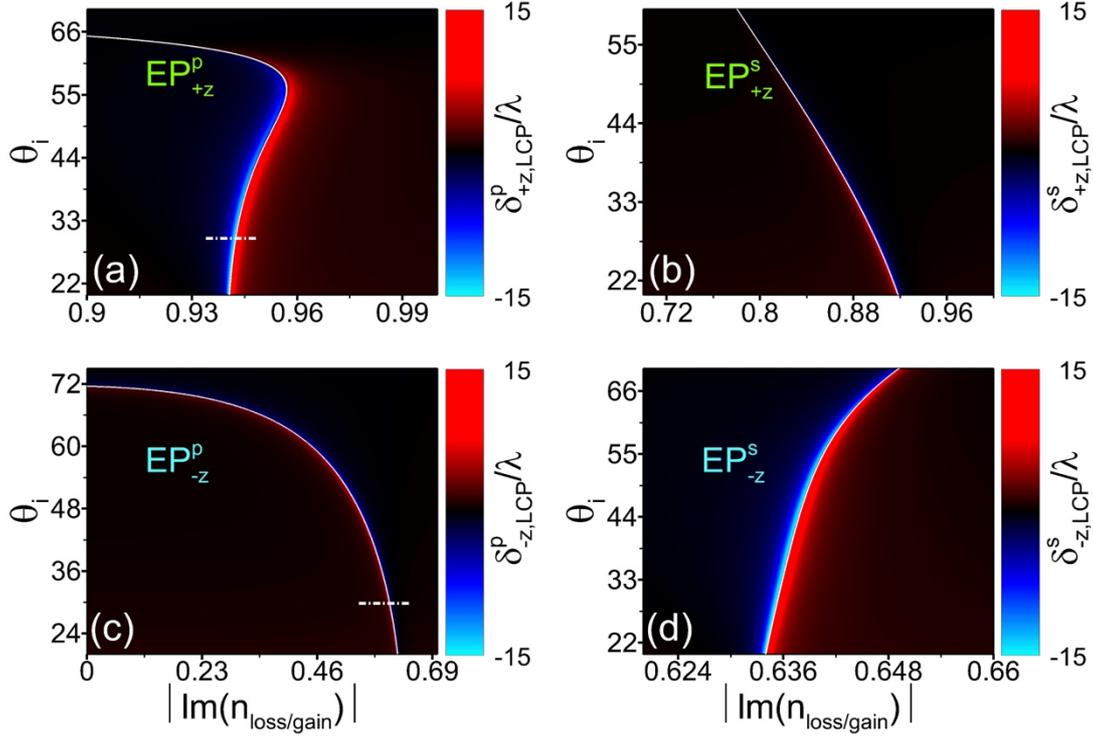

FIG. 2. Representation in parameter space of giant photonic SHE near exceptional points for the reflected beam. Here $\delta_{\text{LCP}}$ is shown in the parameter space, i.e., the space composed by the two independent variables of the imaginary part of refractive index $|Im(n_{\text{loss/gain}})|$ and the incident angle $\theta_i$. The position of exceptional points in the parameter space is dependent on the linear polarization (*p*- or *s*-polarized) and the propagation direction (to the $+z$ or $-z$ direction) of the incident beam. For clarity, these four types of exceptional points are denoted as $\text{EP}^p_{+z}$, $\text{EP}^p_{-z}$, $\text{EP}^s_{+z}$, and $\text{EP}^s_{-z}$, respectively; see the white solid line in each panel. In the parameter space, the PT-symmetric structure has the broken PT-symmetric phase for the region between the two trajectories of $\text{EP}^p_{+z}$ and $\text{EP}^p_{-z}$ for the *p*-polarized incident beam in (a,c) or between the two trajectories of $\text{EP}^s_{+z}$ and $\text{EP}^s_{-z}$ for the *s*-polarized incident beam in (b,d). The PT-symmetric structure has the exact PT phase for the other regions. Since $\delta_{\text{RCP}} = -\delta_{\text{LCP}}$, the value of $\delta_{\text{RCP}}$ can also be directly inferred from this figure. The structural setup is the same as Fig. 1(b). The field distribution of the reflected beam near/at exceptional points is demonstrated in real space in Fig. 4, where the *p*-polarized beam with an incident angle of $\theta_i = 30°$, highlighted by the dashed lines in (a,c), is chosen.

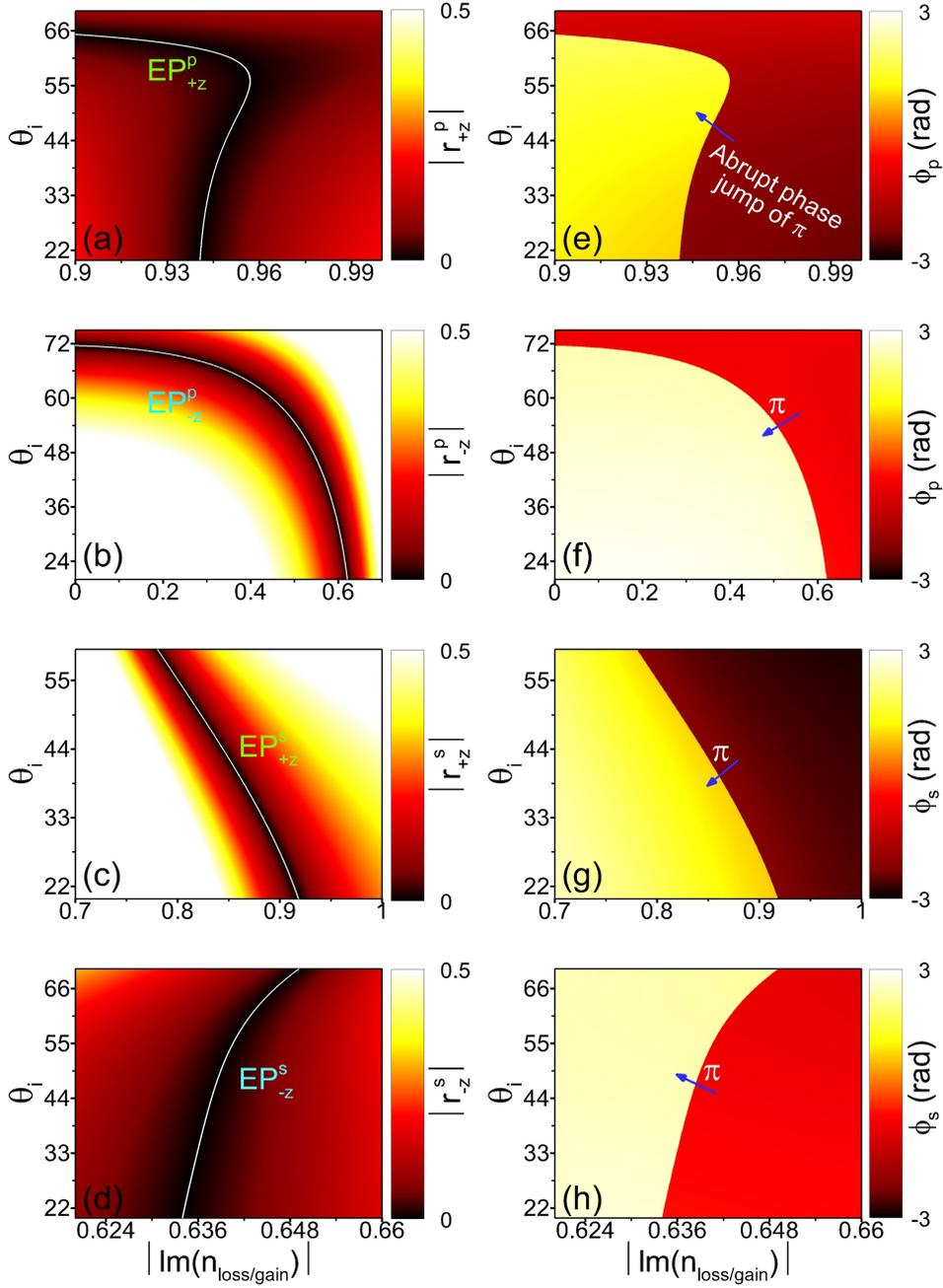

FIG. 3. Explaining the exotic photonic SHE near/at the exceptional points from the perspective of reflection coefficients. At the exceptional points, the reflection coefficients have a zero-value magnitude and an abrupt phase jump of $\pi$. The magnitude and phase of reflection coefficients are shown in (a-d) and (e-h), respectively. The incident beam is *p*-polarized in (a,b,e,f) while is *s*-polarized in (c,d,g,h), and propagates to the $+z$ direction in (a,c,e,g) while propagates to the $-z$ direction in (b,d,f,h). The structural setup is the same as Fig. 1(b).

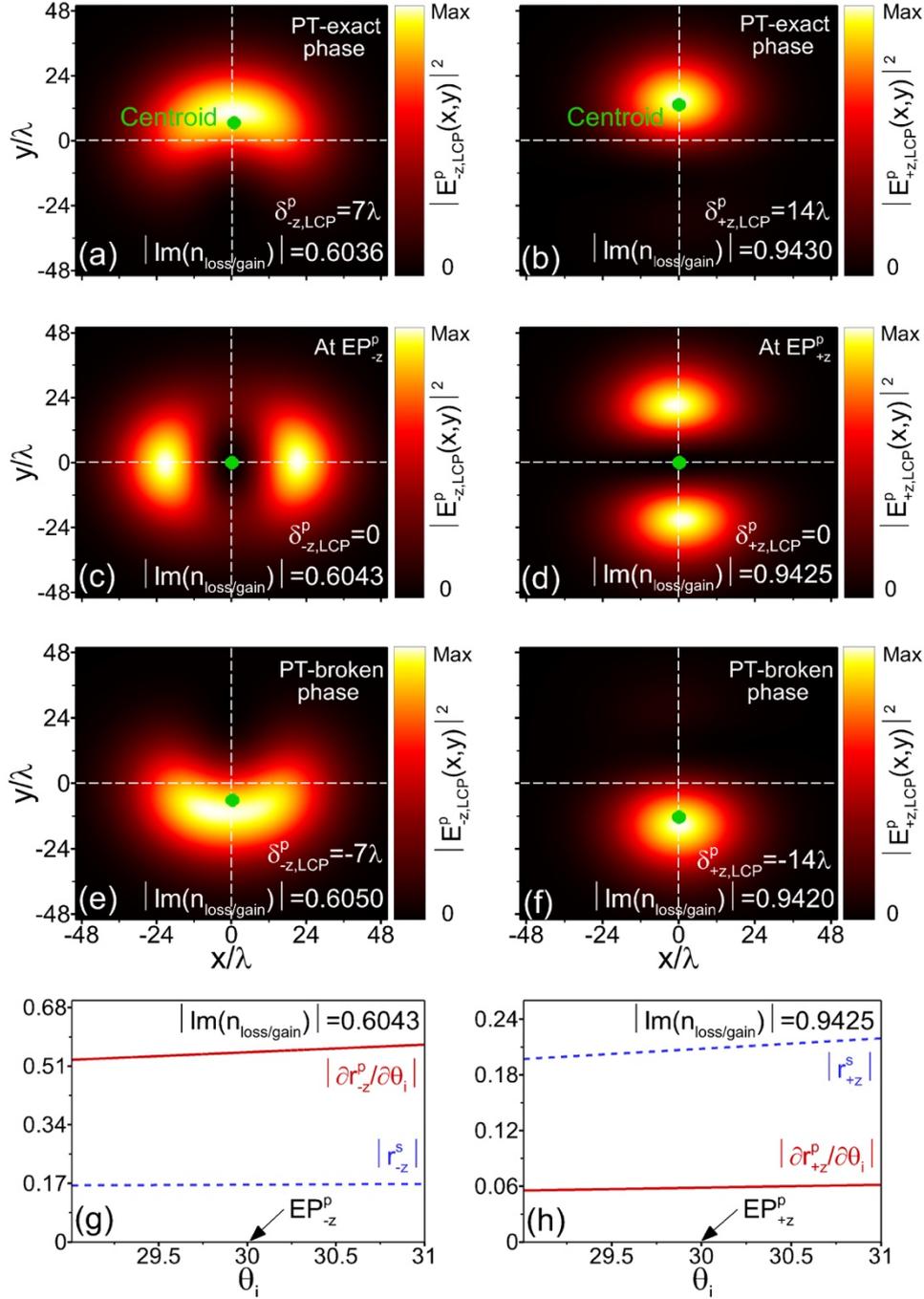

FIG. 4. Real-space demonstration of photonic SHE near/at exceptional points for the reflected beam. (a-f) Field distribution of the reflected beam with LCP at the interface, where the $p$-polarized Gaussian beam with an incident angle of 30° is chosen. For the plots from panel (a), (c), (e), (f), (d) to (b), the value of $|Im(n_{\text{loss/gain}})|$ increases from 0.6036 to 0.9430. At the exceptional points $EP^p_{+z}$ or $EP^p_{-z}$, the transverse centroid of the reflected beam with LCP is at the origin of the $x$-$y$ plane. (g, h) Magnitudes of $|\frac{\partial r^p}{\partial \theta_i}|$ and $|r^s|$ near the exceptional points. The plots in (g, h) can explain the field distribution of the reflected beam at the exceptional points in (c, d).